\documentclass[aps,prl,twocolumn,groupedaddress,showpacs]{revtex4}
\usepackage{graphicx}

\begin{document}

%\preprint{}

\title{Evolution of magnetic polarons and spin-carrier interactions through
the metal-insulator transition in Eu$_{1-x}$Gd$_{x}$O}



%\author{H. Rho,$^{1*}$~C. S. Snow,$^1$ S. L. Cooper,$^1$
%Z. Fisk,$^2$ A. Comment, $^{1,3}$ and J-Ph Ansermet $^{1,3}$}
\author{H. Rho,$^1$ C. S. Snow,$^1$ S. L. Cooper,$^1$
Z. Fisk,$^2$ A. Comment, $^{1,3}$ and J-Ph Ansermet $^{1,3}$}
%\email[Electronic address: ]{rho@mrl.uiuc.edu}
%\author{C. S. Snow,$^1$ S. L. Cooper,$^1$ Z. Fisk,$^2$ A. Comment, $^{1,3}$ and J-Ph Ansermet $^{1,3}$}

%\thanks{}
\affiliation{
$^{1}$Department of Physics and Frederick Seitz Materials Research Laboratory,
University of Illinois at Urbana-Champaign, Urbana, Illinois 61801\\
$^{2}$National High Magnetic Field Laboratory, Florida State University,
1800 East Paul Dirac Drive, Tallahassee, Florida 32306\\
$^{3}$Ecole Polytechnique F\'{e}d\'{e}rale de Lausanne, CH-1015
Lausanne, Switzerland }


\date{\today}

\begin{abstract}

Raman scattering studies as functions of temperature, magnetic
field, and Gd-substitution are used to investigate the evolution
of magnetic polarons and spin-carrier interactions through the
metal-insulator transition in Eu$_{1-x}$Gd$_{x}$O. These studies
reveal a greater richness of phase behavior than have been
previously observed using transport measurements: a
spin-fluctuation-dominated paramagnetic (PM) phase regime for
T\,$>$\,T$^{*}$\,$>$\,T$_{C}$, a two-phase regime for
T\,$<$\,T$^{*}$ in which magnetic polarons develop and coexist
with a remnant of the PM phase, and an inhomogeneous ferromagnetic
phase regime for T\,$<$\,T$_{C}$.

\end{abstract}

%\keywords{}

\maketitle


The remarkable phenomenon of ``colossal magnetoresistance" (CMR)
in perovskite-based oxides such as La$_{1-x}$A$_{x}$MnO$_{3}$,
La$_{2-2x}$A$_{1+2x}$Mn$_{2}$O$_{7}$, and
La$_{1-x}$A$_{x}$CoO$_{3}$ (A\,=\,Sr,\,Ca) has renewed interest in
the general subject of carrier-spin interaction effects in a much
broader class of magnetic semiconductors, including ferromagnets
such as EuO, EuS, and EuB$_{6}$, and antiferromagnets like EuTe
and EuSe \cite{NagaevKasuya}.  In addition to large negative
magnetoresistivities near T$_{C}$, these Eu-based systems share
many intriguing properties with the higher T$_{C}$ perovskite
systems: a metal-insulator (MI) transition, and an accompanying
ferromagnetic-paramagnetic (FM-PM) phase change, near T$_{C}$;
electronic phase-separation tendencies \cite{Nagaev, Moreo};
magnetic-field induced transitions; etc. Consequently, these
binary systems are particularly simple and well-controlled
``laboratories" in which to explore carrier-spin interaction
effects and FM cluster (``magnetic polaron") formation, as well as
the impact of these phenomena on complex magnetic and electronic
phase changes, in a broad class of magnetic systems
\cite{Dagotto}.

The Eu$_{1-x}$Gd$_{x}$O system is particularly interesting to
study for several reasons.  Conductivity and magnetic
susceptibility measurements have shown that Eu-rich EuO exhibits a
FM-PM phase change below T$_{C}$\,$\sim$\,69\,K \cite{Oliver,
Torrance}, affording an opportunity to investigate
spectroscopically the nature of transitions between various
complex phases as functions of temperature and magnetic field.
Furthermore, substitution of Gd provides a means of elevating both
T$_{C}$ \cite{Mauger1} and the conductivity \cite{Oliver} in a
systematic and controlled manner, enabling one to study the
effects of both T$_{C}$ and disorder on spin-carrier interactions
and magnetic polaron formation.

In this Letter, we present an inelastic light (Raman) scattering
study of spin-carrier interactions and magnetic polaron formation
in the Eu$_{1-x}$Gd$_{x}$O system as functions of temperature,
magnetic field, and Gd substitution.  Raman scattering is a
particularly effective technique for studying spin-carrier
interactions in magnetic systems, as it affords a unique means of
simultaneously investigating both the carrier dynamics and spin
excitations in various phases of these materials \cite{Nyhus,
Snow}.  More particularly, spin-flip (SF) Raman scattering
provides a sensitive and direct means by which magnetic polarons
can be detected and studied in different phases of magnetic
semiconductors \cite{Isaacs, Heiman, Peterson, Ramdas, Nyhus,
Snow}.  For example, the Raman scattering study presented here
demonstrates that the phase behavior in Eu$_{1-x}$Gd$_{x}$O is
much richer near T$_{C}$ than has been previously evident from
transport measurements: it consists of a high-temperature PM
regime in which electronic scattering is dominated by spin
fluctuations, a ``cluster formation" temperature regime
(T\,$<$\,T$^{*}$) in which magnetic polarons develop and coexist
with PM regions, and finally an inhomogeneous FM metal regime
(T\,$<\,$T$_{C}$). The results of this study further suggest that
both the stability and exchange energy associated with the
magnetic polarons are enhanced by Gd-substitution and by the
application of a magnetic field.

Raman scattering measurements were performed with
Eu$_{1-x}$Gd$_{x}$O samples (x\,=\,0.006:\,T$_{C}$\,$\sim$\,70\,K;
x\,=\,0.035:\,T$_{C}$\,$\sim$\,115\,K) mounted inside a
variable-temperature, continuous helium-flow cryostat, which
allowed Raman studies at temperatures ranging from 4\,K to 350\,K,
and in magnetic fields up to 8\,T. Samples were excited in a
true-backscattering geometry using the 647.1\,nm excitation
wavelength of a Kr-ion laser. For H\,=\,0 measurements, linearly
polarized light was employed in $z(xx)\bar{z}$ and $z(xy)\bar{z}$
configurations, where $x\|$[1,0,0], $y\|$[0,1,0], and
$z\|$[0,0,1]. In the notation $z(xy)\bar{z}$, $z$ and $\bar{z}$
represent the wavevector directions of the incident and scattered
light, respectively, and $(x,y)$ represents the polarization
directions of the incident and scattered light, respectively.  For
magnetic field dependent measurements, circularly polarized light
was employed in a $z(LR)\bar{z}$ configuration, where $L$ and $R$
represent left and right circularly polarized light, respectively.
The crystallographic axes of each sample were identified by Laue
x-ray diffraction, and the Curie temperatures of the samples were
determined with magnetic susceptibility measurements using
SQUID-based magnetometers.

\begin{figure}
%\centering
%\includegraphics{figure1.eps}
\includegraphics[width=4.9cm]{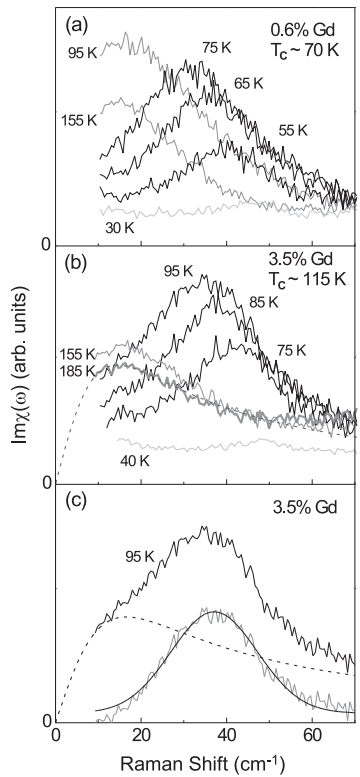}
\vspace{-0.5cm} \caption{\label{}Raman scattering spectra as a
function of temperature for Eu$_{1-x}$Gd$_{x}$O samples with (a)
x\,=\,0.006 and (b) x\,=\,0.035. A dashed line in (b) indicates a
fitting result using Eq.\,1 for the Raman spectrum at 185\,K. The
extended portion of the fitting curve toward 0\,cm$^{-1}$ is a
guide for the eye. (c) SF Raman response (bottom gray curve),
which has been fit with a Gaussian profile (dark solid line),
after removing collision-dominated contribution (dashed line) from
the raw T\,=\,95\,K spectrum (top curve).} \vspace{-0.6cm}
\end{figure}

Conductivity measurements of EuO \cite{Torrance, Oliver} can
distinguish two distinct temperature regimes in this system: a
regime above T$_{C}$ ($\sim$\,69\,K) in which the resistivity
exhibits activated behavior and a large negative
magnetoresistivity, and a regime below T$_{C}$ in which the
conductivity increases by roughly 13 orders-of-magnitude as the
system transitions into the FM metal phase.  By contrast, as shown
in Figs.\,1\,(a) and (b), Raman scattering measurements are able
to resolve additional richness in the phase behavior of this
system, revealing distinct spectroscopic signatures associated
with three different phase regimes as a function of temperature:
(i) a high temperature ``spin-fluctuation-dominated" PM regime for
T$_{C}$\,$<$\,T\,$<$\,T$_{sf}$, where T$_{sf}$ is the temperature
below which carrier scattering is dominated by spin fluctuations,
(ii) a ``magnetic polaron" regime in which magnetic polarons
develop below a formation temperature T$^*$
(T$_{C}$\,$<$\,T$^*$\,$<$\,T$_{sf}$), and (iii) an inhomogeneous
FM regime for T\,$<$\,T$_{C}$ in which a FM metal component
coexists with a diminishing density of magnetic polarons.  Below,
we consider each of these phase regimes, and the transitions
between them, in greater detail.

The high temperature PM phase is characterized in the Raman
spectrum by a collision-dominated electronic scattering response
\cite{Zawadowski, Nyhus, Snow},
\begin{equation}
S(\omega)\propto(1+n(\omega))Im\chi(\omega)=(1+n(\omega))
\frac{|\gamma_{L}|^2\omega\Gamma_L}{\omega^2+\Gamma_{L}^{2}},
\end{equation}
where $1+n(\omega)$ is the Bose thermal factor, $L$ is the
scattering channel selected by a particular scattering geometry,
$\gamma_{L}$ is the Raman scattering vertex in channel L, and
$\Gamma_{L}$ is the electronic scattering rate in channel L.
Fig.\,1\,(b) illustrates that the collision-dominated response in
Eq.\,1 indeed provides an excellent fit (dashed line) to the Raman
spectra in the PM phase. In conventional semiconductors, the
scattering rate is typically associated with impurity scattering
due to extrinsic defects or vacancies \cite{Klein}.  However, in
magnetic semiconductors such as EuO and EuB$_{6}$, for
temperatures sufficiently close to T$_C$, electronic scattering is
dominated by short-range spin fluctuations.  Consequently, we find
in this regime that the scattering rate scales according to
$\Gamma$\,$\sim$\,$\chi(T)T$ \cite{Snow}, where $\chi(T)$ is the
temperature-dependent magnetic susceptibility, and we also find
that the intensity of the collision-dominated scattering response
increases substantially with decreasing temperature towards T$_C$.
Significantly, the resistivity of Eu$_{1-x}$Gd$_{x}$O
\cite{MolnarJAP} exhibits activated behavior between room
temperature and T$_C$, which also suggests the dominance of
``critical" electronic scattering from spin fluctuations
throughout this temperature regime.  However, the Raman results in
Fig.\,1\,(a) and (b) reveal additional complexity in the H\,=\,0
phase behavior of Eu$_{1-x}$Gd$_{x}$O above T$_C$: below a
temperature T$^{*}$ ($>$\,T$_C$), there is a striking change in
the Raman response, from a collision-dominated low frequency
response to an inelastic response with a clear Gaussian profile.
Significantly, this inelastic response develops in the
(\emph{x},\,\emph{y}) and (\emph{x+y},\,\emph{x-y}) [i.e.,
E$_i$\,$\perp$\,E$_s$] scattering geometries, but not in the
(\emph{x},\,\emph{x}) [i.e., E$_i$\,$\|$\,E$_s$] scattering
geometry, and hence has the transformation properties of the
totally antisymmetric Raman tensor. Numerous previous
investigations in both dilute and dense magnetic semiconductors
have identified this distinctive response as H\,=\,0 SF Raman
scattering associated with the development of magnetic polarons
\cite{Isaacs, Heiman, Peterson, Ramdas, Nyhus, Snow}. Hence, the
development of this Raman response in the Eu$_{1-x}$Gd$_{x}$O
system betrays a distinct regime above T$_C$ in which local FM
clusters nucleate prior to the development of the FM ground state.
As a function of decreasing temperature below T$_C$, the H\,=\,0
SF Raman response gradually decreases in intensity as the system
transitions into the FM metal phase, reflecting the gradual
dissolution of localized magnetic polarons with increasing spin
order in the FM phase as the localization length for the magnetic
polarons diverges.  Below 50\,K, the Raman intensity associated
with magnetic polarons decreases rapidly due to the saturation of
spins.

\begin{figure}
%\centering
\includegraphics[width=5cm]{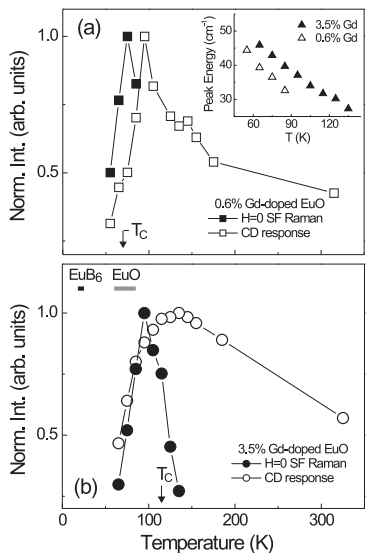}
\vspace{-0.4cm} \caption{\label{}Spectrally integrated intensity
changes as a function of temperature both for collision-dominated
scattering response and for magnetic polaron response in
Eu$_{1-x}$Gd$_{x}$O with (a) x\,=\,0.006 and (b) x\,=\,0.035.
Horizontal bars in (b) denote temperature ranges in which magnetic
polarons form in EuO and EuB$_6$ for a comparison purpose.  The
inset in (a) shows polaron peak energy changes as a function of
temperature for x\,=\,0.006 and 0.035.} \vspace{-0.6cm}
\end{figure}

Figures 1 and 2 reveal several important features of magnetic
polaron evolution in Eu$_{1-x}$Gd$_{x}$O.  First, the H\,=\,0 SF
Raman energy increases systematically with decreasing temperature,
indicative of magnetic polarons in a spin-aligned ``cooperative"
regime, in which the spins of the carriers and magnetic ions are
cooperatively aligned in the FM clusters \cite{Heiman, Isaacs,
Warnock}.  Second, careful fits to the spectra, illustrated in
Fig.\,1\,(c) and summarized in Fig.\,2, demonstrate that the
magnetic polaron regime is characterized by the coexistence of
H\,=\,0 SF (inelastic Gaussian) and collision-dominated electronic
(Eq.\,1) scattering responses, providing strong evidence that this
is a ``two-phase" regime in which FM clusters coexist with some
remnant of the PM phase.  This behavior is summarized for both
x\,=\,0.006 and x\,=\,0.035 samples in Fig.\,2, which compares the
integrated intensities of both the H\,=\,0 SF (filled symbols) and
collision-dominated scattering (open symbols) responses as a
function of temperature.  Finally, note that the polarons in
Eu$_{1-x}$Gd$_{x}$O are stable over a temperature range that is
$\sim$5--10 times higher than in EuB$_6$ \cite{Nyhus, Snow}.
Furthermore, both $T^{*}$ and the temperature range over which
polarons are stable increase with increasing Gd-concentration.
These results provide direct evidence that increased spin-disorder
stabilizes magnetic polarons in the CMR-type systems.

\begin{figure}
%\centering
\includegraphics[width=5.3cm]{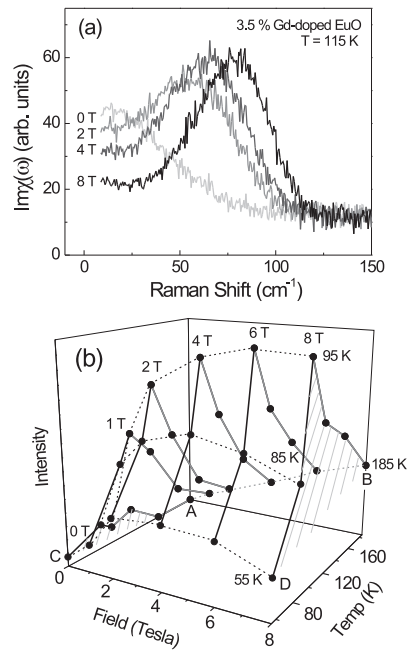}
\vspace{-0.3cm} \caption{\label{}(a) Magnetic field dependent
Raman scattering at 115\,K.  Note that in a $z(LR)\bar{z}$
scattering configuration, the collision-dominated scattering
response prevails in spectra when H\,$<$\,1\,T, that makes unable
to identify more detail on the SF Raman response. (b) Spectrally
integrated intensity profile of the SF Raman response as functions
of magnetic field and temperature. Data points observed at the
same field (temperature) are connected by the solid (dotted)
lines.} \vspace{-0.6cm}
\end{figure}

It is of great interest to examine the effects of an externally
applied field on the low frequency excitation spectra near T$_C$
in Eu-based compounds, and more particularly on the magnetic
polarons in these systems.  The field dependence of the Raman
spectrum is illustrated for T\,=\,115\,K in Fig.\,3\,(a), which
shows the evolution of a SF Raman response with increasing
magnetic field. The integrated intensity of the SF Raman response
is summarized as functions of both the temperature and magnetic
field in Fig.\,3\,(b).  Figures\,3\,(a) and (b) show two
particularly interesting effects of the magnetic field on the SF
Raman spectrum in Eu$_{0.965}$Gd$_{0.035}$O.  First, both the
integrated intensity and the energy of the SF Raman response
increase with increasing field, suggesting that there is a
corresponding increase in the effective d-f exchange energy within
the magnetic cluster. Second, the collision-dominated electronic
response diminishes with increasing field, reflecting the decrease
of spin-fluctuations at high fields.  The SF Raman intensity
``surface" shown in Fig.\,3\,(b) also reveals several features:
(a) in the PM phase, increasing the magnetic field from H\,=\,0
(i.e., from point A to B at T\,=\,185\,K) results in the
development and linear increase in the SF Raman intensity,
suggesting that a magnetic field stabilizes the formation of
magnetic polarons at high temperature, presumably by both
increasing the magnetic susceptibility at these temperatures and
reducing thermal fluctuations; and (b) increasing the magnetic
field from H\,=\,0 well below T$_C$ (e.g., from point C to D at
T\,=\,55\,K) causes the SF Raman intensity to increase initially
at lower fields, then decrease for higher fields, suggesting that
the system is close to the FM phase regime in which the polarons
become ionized at these high fields and low temperatures.  This
trend is similar to that observed in EuB$_6$ at high fields
\cite{Nyhus, Snow}.

\begin{figure}
%\centering
\vspace{-0.5cm}
\includegraphics[width=5.5cm]{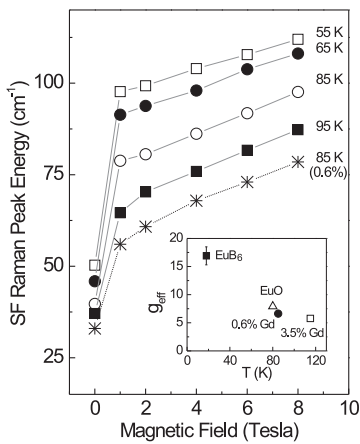}
\vspace{-0.6cm} \caption{\label{}SF Raman peak energy changes as a
function of applied magnetic field at various temperatures for
Eu$_{1-x}$Gd$_{x}$O with x\,=\,0.006 (stars) 0.035 (squares and
circles).  The inset shows effective g-value obtained from linear
fits in the field range of 1\,$\leq$\,H\,$\leq$\,8\,T.}
\vspace{-0.7cm}
\end{figure}

In order to examine more quantitatively the effects of a magnetic
field on the magnetic polarons in Eu$_{1-x}$Gd$_{x}$O, we plot in
Fig.\,4 the SF Raman energy as a function of field at several
temperatures, and for both 0.6$\%$ Gd- (stars) and 3.5$\%$ Gd-
(squares and circles) substituted EuO.  To simplify the
interpretation of these data, we consider only field-dependent
data at those temperatures for which there is an H\,=\,0 SF Raman
response, i.e., for which there are magnetic polarons present even
at H\,=\,0. The summary in Fig.\,4 illustrates several interesting
points. First, there is an abrupt jump in the SF Raman peak energy
between 0--1 Tesla, followed by a linear increase in the SF energy
with increasing magnetic field. As the SF Raman energy is given by
$\hbar$$\omega_{0}$\,$\propto$\,$J$$_{df}$$\langle$$M$$\rangle$,
the increase with increasing field reflects an enhancement in the
exchange energy associated with the polarons due to increased
polarization of the spins.  The origin of the abrupt jump with
small applied field is not clear, but it is worth noting that the
individual moments of the spontaneously-formed magnetic polarons
are randomly oriented at H\,=\,0, but are expected to become
rapidly aligned upon application of a field \cite{Dagotto}. Hence,
the abrupt jump in the SF Raman energy appears to reflect an
enhancement in the effective field experienced by localized
charges when the magnetic polarons become mutually-aligned with
the application of a field.  Second, Fig.\,4 illustrates that the
SF Raman energy increases both with decreasing temperature and
with increasing Gd concentration, again presumably reflecting the
larger effective magnetization associated with the polarons (i.e.,
larger polaron sizes) under these conditions \cite{Dagotto,
Guillaume}. Using the slope obtained from the linear-in-field
``high-field" regime (1\,$\leq$\,H\,$\leq$\,8\,T), we can estimate
an effective g-value
($g_{eff}$\,$\sim$\,$\Delta$E/$\mu_B$$\Delta$H) of $\sim$\,6 for
0.6$\%$ and 3.5$\%$ (Eu,Gd)O at T\,$\sim$\,T$_C$.  By contrast,
the much less disordered system EuB$_6$ has g$_{eff}$\,$\sim$\,17
for T\,$\sim$\,T$_C$, suggesting that the polaron size in (Eu,Gd)O
may be fundamentally much more limited by intrinsic spin disorder
than in EuB$_6$.  Indeed, the PM semimetal to FM metal transition
in EuB$_6$ appears to occur via a continuous evolution --- and
eventual percolation --- of the magnetic polarons \cite{Nyhus,
Snow}, whereas the polaronic SF Raman response in (Eu,Gd)O simply
appears to dissipate into the FM phase (see Fig.\,3\,(b)) with
decreasing temperature.

In summary, Raman scattering studies reveal direct evidence for
diverse phase behavior in Eu$_{1-x}$Gd$_{x}$O, most notably a
coexistence regime involving magnetic polarons and PM phase
regions in the vicinity of T$_C$.  This regime is found to be
stabilized both by increasing magnetic field and the substitution
of magnetic impurities.  These results demonstrate, most
significantly, that electronic inhomogeneity and cluster formation
are not unique to complex, perovskite-related oxides
\cite{Dagotto}, but occur rather generally even in structurally
simple (binary) systems that exhibit a similar strong competition
between carrier kinetic and magnetic interaction energies.

We thank M. V. Klein for useful discussions.  We acknowledge
support of this work by the Department of Energy
(DEFG02-96ER45439) and the National Science Foundation
(DMR97-00716).

\vspace{-0.7cm}
\bibliography{basename of .bib file}

\begin{thebibliography}{}

%\bibitem[*]{email} rho@mrl.uiuc.edu
\vspace{-0.7cm}

\bibitem{NagaevKasuya}For a review of these materials, see E. L.
Nagaev, Physics Reports, \textbf{346}, 387 (2001) and T. Kasuya
and A. Yanase, Rev. Mod. Phys. \textbf{40}, 684 (1968).

\bibitem{Nagaev} E. L. Nagaev, Phys. Stat. Sol. B \textbf{186}, 9 (1994).

\bibitem{Moreo} A. Moreo \emph{et al.},
Phys. Rev. Lett. \textbf{84}, 5568 (2000); A. Moreo \emph{et al.},
Science \textbf{283}, 2034 (1999).

\bibitem{Dagotto} J. Burgy \emph{et al.},
arXiv:cond-mat/0107300 (2001).

\bibitem{Oliver} M. R. Oliver \emph{et al.},
Phys. Rev. B \textbf{5}, 1078 (1972).

\bibitem{Torrance} J. B. Torrance \emph{et al.}, Phys. Rev. Lett.
\textbf{29}, 1168 (1972).

\bibitem{Mauger1} A. Mauger \emph{et al.},
J. Phys. (Paris) \textbf{41}, C5-263 (1980).

\bibitem{Nyhus} P. Nyhus \emph{et al.}, Phys. Rev. B \textbf{56}, 2717 (1997).

\bibitem{Snow} C. S. Snow \emph{et al.},
to be published in Phys. Rev. B.

\bibitem{Isaacs} E. D. Isaacs \emph{et al.},
Phys. Rev. B \textbf{37}, 7108 (1988).

\bibitem{Heiman} D. Heiman \emph{et al.}, Phys. Rev. B \textbf{27}, 4848 (1983).

\bibitem{Peterson} D. L. Peterson \emph{et al.}, Phys. Rev. B \textbf{32}, 323 (1985).

\bibitem{Ramdas}  A. K. Ramdas and S. Rodriguez, in {\it Light Scattering in Solids VI},
edited by M. Cardona and G. G\"untherodt (Springer-Verlag, Berlin,
1991), p. 137.

\bibitem{Zawadowski} A. Zawadowski and M. Cardona, Phys. Rev. B \textbf{42}, 10732 (1990).

\bibitem{Klein}  M. V. Klein, in {\it Light Scattering in Solids}, edited by M. Cardona
(Springer-Verlag, Berlin, 1975), p. 150.

\bibitem{MolnarJAP} S. von Molnar and M. W. Shafer, J. Appl. Phys. \textbf{41}, 1093 (1970).

\bibitem{Warnock} J. Warnock and P. A. Wolff, Phys. Rev. B \textbf{31}, 6579 (1985).

\bibitem{Guillaume} C. B. $\grave{a}$ la Guillaume, Physica \textbf{146B}, 234 (1987).


%\bibitem{NagaevKasuya}For a review of these materials, see E. L.
%Nagaev, Physics Reports, \textbf{346}, 387 (2001) and T. Kasuya
%and A. Yanase, Rev. Mod. Phys. \textbf{40}, 684 (1968).
%\bibitem{Nagaev} E. L. Nagaev, Phys. Stat. Sol. B \textbf{186}, 9 (1994).
%\bibitem{Moreo} A. Moreo, M. Mayr, A. Feiguin, S. Yunoki, and E. Dagotto,
%Phys. Rev. Lett. \textbf{84}, 5568 (2000); A. Moreo, S. Yunoki, E.
%Dagotto, Science \textbf{283}, 2034 (1999).
%\bibitem{Dagotto} J. Burgy, M. Mayr, V. Martin-Mayor, A. Moreo, and E. Dagotto,
%arXiv:cond-mat/0107300 (2001).
%\bibitem{Oliver} M. R. Oliver, J. O. Dimmock, A. L. McWhorter, and T. B. Reed,
%Phys. Rev. B \textbf{5}, 1078 (1972).
%\bibitem{Torrance} J. B. Torrance, M. W. Shafer, and T. R. McGuire, Phys. Rev. Lett.
%\textbf{29}, 1168 (1972).
%\bibitem{Mauger1} A. Mauger, M. Escorne, C. Godart, J. P. Desfours, and J. C. Achard,
%J. Phys. (Paris) \textbf{41}, C5-263 (1980).
%\bibitem{Nyhus} P. Nyhus, S. Yoon, M. Kauffman, S. L. Cooper, Z.
%Fisk, and J. Sarrao, Phys. Rev. B \textbf{56}, 2717 (1997).
%\bibitem{Snow} C. S. Snow, S. L. Cooper, D. P. Young, Z. Fisk, A. C. Comment, and J-Ph. Ansermet,
%to be published in Phys. Rev. B.
%\bibitem{Isaacs} E. D. Isaacs, D. Heiman, M. J. Graf, B. B. Goldberg, R. Kershaw,
%D. Ridgley, K. Dwight, A. Wold, J. Furdyna, and J. S. Brooks,
%Phys. Rev. B \textbf{37}, 7108 (1988).
%\bibitem{Heiman} D. Heiman, P. A. Wolff, and J. Warnock, Phys. Rev. B \textbf{27}, 4848 (1983).
%\bibitem{Peterson} D. L. Peterson, D. U. Bartholomew, U. Debska, A. K. Ramdas, and
%S. Rodriguez, Phys. Rev. B \textbf{32}, 323 (1985).
%\bibitem{Ramdas}  A. K. Ramdas and S. Rodriguez, in {\it Light Scattering in Solids VI},
%edited by M. Cardona and G. G\"untherodt (Springer-Verlag, Berlin,
%1991), p. 137.
%\bibitem{Zawadowski} A. Zawadowski and M. Cardona, Phys. Rev. B \textbf{42}, 10732 (1990).
%\bibitem{Klein}  M. V. Klein, in {\it Light Scattering in Solids}, edited by M. Cardona
%(Springer-Verlag, Berlin, 1975), p. 150.
%\bibitem{MolnarJAP} S. von Molnar and M. W. Shafer, J. Appl. Phys. \textbf{41}, 1093 (1970).
%\bibitem{Warnock} J. Warnock and P. A. Wolff, Phys. Rev. B \textbf{31}, 6579 (1985).
%\bibitem{Guillaume} C. B. $\grave{a}$ la Guillaume, Physica \textbf{146B}, 234 (1987).

\end{thebibliography}

\end{document}